\documentclass[conference]{IEEEtran}
\usepackage{cite}
\usepackage{amsmath,amssymb,amsfonts}
\usepackage{algorithmic}
\usepackage{graphicx}
\usepackage{textcomp}
\usepackage{xcolor}
\usepackage[hyphens]{url}
\usepackage[version=4]{mhchem}
\usepackage{tabularx}
\usepackage{booktabs}

\usepackage{caption}
\usepackage{subfigure}

\graphicspath{{fig/}}

\def\BibTeX{{\rm B\kern-.05em{\sc i\kern-.025em b}\kern-.08em
    T\kern-.1667em\lower.7ex\hbox{E}\kern-.125emX}}

\IEEEoverridecommandlockouts

\title{SMILES-Inspired Transfer Learning for Quantum Operators in Generative Quantum Eigensolver}
\author{
\IEEEauthorblockN{Zhi Yin\IEEEauthorrefmark{1}\IEEEauthorrefmark{3}\thanks{Corresponding author: Zhi Yin (yinzhi@quregenai.com, yz@nbut.edu.cn).},
Xiaoran Li\IEEEauthorrefmark{1}\IEEEauthorrefmark{2},
Shengyu Zhang\IEEEauthorrefmark{4},
Xin Li \IEEEauthorrefmark{4},
Xiaojin Zhang \IEEEauthorrefmark{1}\IEEEauthorrefmark{2}}
\IEEEauthorblockA{\IEEEauthorrefmark{1}
QureGenAI(AceMapAI) Biotechnology, Suzhou, 215000, China}
\IEEEauthorblockA{\IEEEauthorrefmark{2}QureGenAI(AceMapAI) Joint Lab, China Pharmaceutical University, Nanjing, 211198, China}
\IEEEauthorblockA{\IEEEauthorrefmark{3}School of Statistics and Data Science, Ningbo University of Technology, Ningbo, 315211, China}
\IEEEauthorblockA{\IEEEauthorrefmark{4}Tencent Quantum Lab, Shenzhen, 518057, China}
}

\begin{document}
\maketitle
\thispagestyle{plain}
\pagestyle{plain}


\begin{abstract}

Given the inherent limitations of traditional Variational Quantum Eigensolver(VQE) algorithms, the integration of deep generative models into hybrid quantum-classical frameworks, specifically the Generative Quantum Eigensolver(GQE), represents a promising innovative approach. However, taking the Unitary Coupled Cluster with Singles and Doubles(UCCSD) ansatz which is widely used in quantum chemistry as an example, different molecular systems require constructions of distinct quantum operators. Considering the similarity of different molecules, the construction of quantum operators utilizing the similarity can reduce the computational cost significantly. 
Inspired by the SMILES representation method in computational chemistry, we developed a text-based representation approach for UCCSD quantum operators by leveraging the inherent representational similarities between different molecular systems. This framework explores text pattern similarities in quantum operators and employs text similarity metrics to establish a transfer learning framework. Our approach with a naive baseline setting demonstrates knowledge transfer between different molecular systems for ground-state energy calculations within the GQE paradigm. This discovery offers significant benefits for hybrid quantum-classical computation of molecular ground-state energies, substantially reducing computational resource requirements. 

\end{abstract}

\section{Introduction}

Hybrid quantum-classical algorithms~\cite{cerezo2021variational,bharti2022noisy}, exemplified by Variational Quantum Eigensolver(VQE)~\cite{peruzzo2014variational,grimsley2019adaptive} have demonstrated significant promise for addressing computational challenges in molecular electronic ground state energy calculation problem. Since VQE approaches have been plagued by the Barren Plateau(BP)~\cite{magann2023randomized,wang2021noise} problem, characterized by exponentially vanishing gradients in high-dimensional parameter spaces and scalability challenges arising from exponential resource scaling with system complexity in molecular systems, a novel framework termed the Generative Quantum Eigensolver(GQE)~\cite{nakaji2024generative,minami2025generative} has been proposed to address the challenges of barren plateaus and the issue of scalability limitation in VQE. Unlike traditional VQE that relies on gradient-based optimization of parametrized quantum circuits, GQE uses generative models to produce sequences of quantum operators that progressively reduce the energy expectation value. This new methodology circumvents the optimization of parametrized quantum circuits in conventional VQE by constructing classical generative model from a pre-defined quantum operators pool and samples to generate desired circuit architectures. 
GQE offers several key advantages, including improved exploration of the solution space, reduced susceptibility to barren plateau, and enable resource-efficient quantum circuit ansatz design without explicit parameter tuning. Moreover, GQE demonstrates remarkable adaptability to the molecular electronic ground state energy calculation \cite{nakaji2024generative}.


Despite these advantages, current GQE implementations face significant efficiency challenges that limit their practical applicability. First, GQE models typically require training from scratch for each new molecular system, failing to leverage the substantial structural and electronic similarities between related molecules. This approach requires repeated resource-intensive training cycles, even for molecules with similar characteristics, resulting in significant computational redundancy. Furthermore, as molecular complexity increases, the dimensionality of the quantum operator pool grows combinatorially, making the learning problem increasingly difficult. Consider the Unitary Coupled Cluster Singles and Doubles(UCCSD)~\cite{barkoutsos2018quantum} ansatz, which is widely regarded as representations for quantum chemistry calculations and generates quantum operators whose size scales polynomially with the number of molecular orbitals. Although GQE supports transfer learning, effective molecule-to-molecule(mol-to-mol) transfer across arbitrary small molecules has not been adequately demonstrated. This limits their practical utility for real-world applications in computational chemistry. To address these limitations, we propose a novel approach inspired by the Simplified Molecular Input Line Entry System(SMILES)~\cite{weininger1988smiles} representation widely used in cheminformatics. Drawing on this analogy, we recognize that quantum operators generated by UCCSD can similarly be represented as text patterns, with inherent similarities that can be exploited through transfer learning~\cite{weiss2016survey,torrey2010transfer,zhuang2020comprehensive}. 


Our framework treats quantum operators as structured textual representations with inherent patterns and similarities across different molecular systems. By quantifying the text-based similarity between operators, we establish mappings between the source and target molecules that preserve functional relationships while accommodating differences in the dimensionality of GQE model. This approach enables knowledge transfer from previously trained GQE models to new molecular systems, significantly reducing the computational resources required for ground state energy predictions. The transfer learning process involves adapting embedding layers and output heads while preserving the core transformer architecture that has captured generalizable patterns in quantum operators sequences. Our approach bridges the gap between the source and target molecules while maintaining computational efficiency.

Our preliminary experiments demonstrate the viability of our approach even with a naive baseline setting. Using simple sequence matching for operator similarity and limited fine-tuning, we observed discernible patterns in transfer performance that correlate with molecular similarity. Furthermore, certain distinct trends were identified within our transfer learning paradigm, and this suggests potential for improvement with more sophisticated representation learning techniques. Moreover, our method demonstrates remarkable computational efficiency compared to the training from scratch GQE, enabling rapid molecular ground-state energy calculations in future application.

In the following sections, we detail our methodology, present experimental results across diverse molecular systems, summarize our proposed approach, and outline potential directions for future improvements.

\section{Methods}

Our methodology addresses the challenge of efficiently computing molecular ground state energies by leveraging similarities in quantum operator representations across different molecular systems within GQE framework. The approach combines the generative capabilities of quantum eigensolver with transfer learning techniques, enabling knowledge transfer from source to target molecules. 

UCCSD is a prominent approach for mapping molecular electronic-structure information into a parameterized quantum circuit. Rooted in quantum chemistry, it possesses clear physical meaning and effectively captures electron correlation under standard assumptions. Accordingly, UCCSD can be viewed as a mechanism that maps molecular electronic structure into a quantum-circuit space, where the circuit is composed of a sequence of quantum operators.

In molecular representation, SMILES is widely adopted for encoding two-dimensional structural information as strings. For example, ring openings and closures are marked by matching digits, chemical bonds are denoted by symbols (single bonds by default omission, \textbf{=} for double bonds, \textbf{\#} for triple bonds), and stereochemical configurations are indicated by \textbf{@} and \textbf{@@}. This representation is highly flexible and semantically expressive, enabling compositional, language-like encodings of diverse molecular structures.

Quantum operators admit multiple, complementary representations, each exposing different structural and computational properties. (i) Graph-based representations: Circuit or operator-level graphs (e.g. directed acyclic graphs) capture causal structure, dependencies, and transformation rules, facilitating optimization and scheduling. (ii) Exponential/series expansions: Operators expressed via exponential maps and their series. (iii) Matrix and tensor forms: Dense or structured matrices as well as tensor-network factorizations. (iv) Textual/serialized encodings: Human-readable or machine-serializable formats like SMILES provided by standard quantum computing libraries allow interoperable specification of operators and circuits.

Although representations differ in expressive capacity, by analogy we choose SMILES as the baseline because it is the simplest and interpretable option. We propose to represent UCCSD circuits corresponding to different molecular electronic structures as strings in the spirit of SMILES. In this scheme, single and double excitation operators serve as the fundamental circuit units. At a finer granularity, each operator can be decomposed into labeled fields specifying (i) excitation type (single vs. double), (ii) the rotation angle(s) of the controlled operations, and (iii) the indices of the acted-on qubits. Given these elements, one can systematically assemble the corresponding UCCSD circuit.

In summary, molecular electronic structures, when transformed by UCCSD, can be encoded as SMILES-like strings. Owing to UCCSD’s firm grounding in quantum-chemical principles, this encoding expresses electronic structure as combinations of well-defined quantum operators, thereby enabling efficient representation, comparison, and exploration of distinct electronic structures within the resulting representation hidden space.

The following section will elaborate on our representation framework and the associated transfer learning pipeline. It is worth emphasizing that this rudimentary baseline setting is purposefully constrained to establish methodological feasibility and provide a benchmark for subsequent refinements.


\subsection{Representation of UCCSD Operators}

The foundation of our method lies in the text-based representation of quantum operators. We utilize the UCCSD ansatz, which generates operators consisting of single and double excitation operations of specific molecules. These operations form the building blocks of quantum circuits designed to approximate molecular ground states. In basic GQE approaches, each molecular system requires constructing and optimizing a specific operator pool, leading to computational redundancy for similar molecules.

\begin{figure}[htbp]
    \centering
    \includegraphics[width=0.8\linewidth]{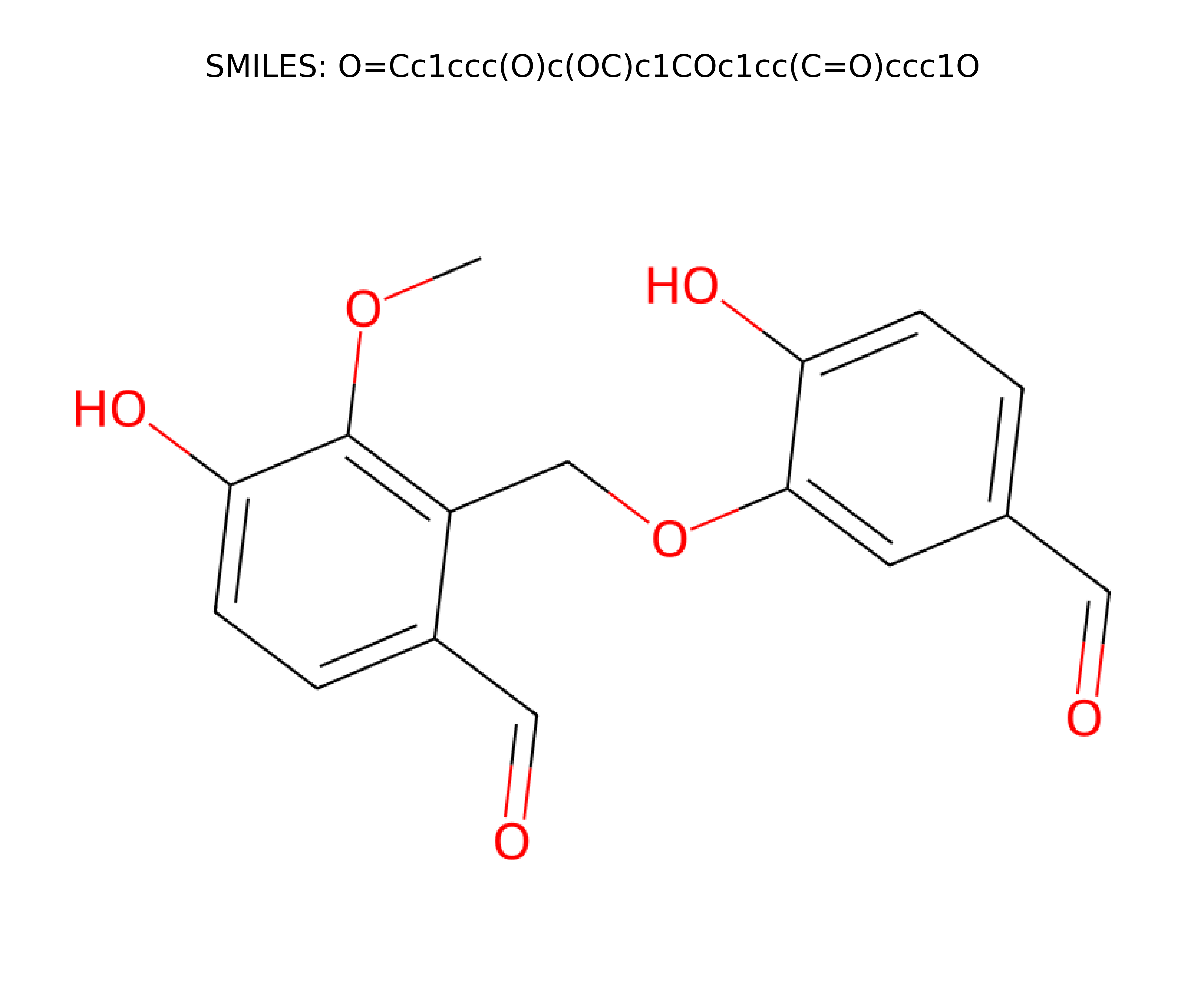}  
    \caption{A representative example of a molecule along with its SMILES notation.}  
    \label{fig:smiles}  
\end{figure}

\begin{figure}[htbp]
    \centering
    \includegraphics[width=0.7\linewidth]{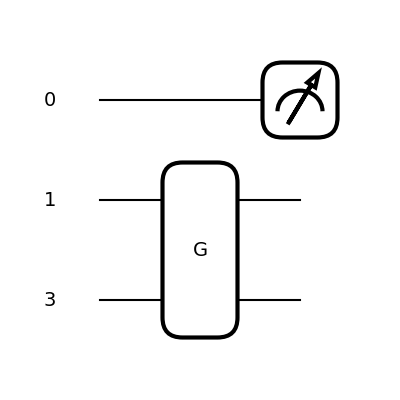}  
    \caption{Demonstration of a typical quantum operation with its corresponding string representation: \texttt{qml.SingleExcitation(-0.05, wires=[1, 3])}.}  
    \label{fig:circuit}  
\end{figure}

Our representation approach, inspired by the SMILES notation (Fig.~\ref{fig:smiles}) in computational chemistry, treats quantum operators as textual patterns. Using the PennyLane quantum computing library~\cite{bergholm2018pennylane}, operators are simply expressed as string representations such as: \texttt{qml.SingleExcitation(-0.1, wires=[1, 3])} (Fig. ~\ref{fig:circuit}). 

These string representations capture essential information about each quantum operation, including:

\begin{enumerate}
\item The operation type (e.g., SingleExcitation, DoubleExcitation).
\item The parameter value controlling operation.
\item The specific qubits (wires) on which the operation acts
\end{enumerate}

Just as SMILES strings enable efficient molecular representation for machine learning in chemistry, our approach enables quantum operators to be treated as tokens in a vocabulary, allowing language model techniques~\cite{radford2019language,vaswani2017attention,zhao2021point,dosovitskiy2020image} to be applied to quantum circuit generation~\cite{furrutter2024quantum,daimon2024quantum}.


\subsection{Similarity Measurement and Mapping}
To establish connections between quantum operators in source and target molecules, we implement a similarity measurement framework. In our baseline implementation, we utilize string similarity metrics to determine correspondences between operators. Specifically, we employ sequence matching techniques that quantify the degree of similarity between string representations of quantum operators.

\begin{figure}[htbp]
    \centering
    \includegraphics[width=1\linewidth]{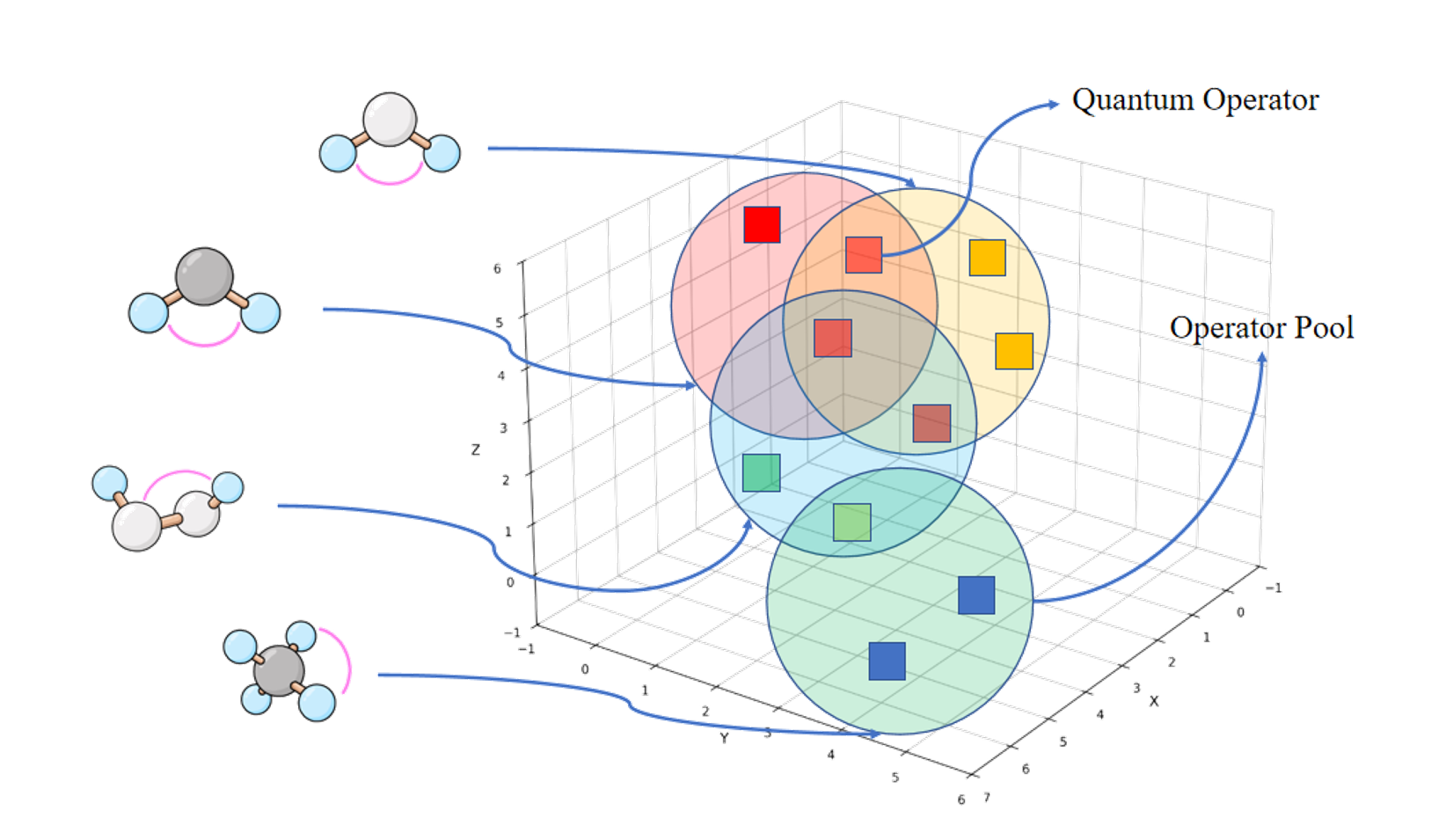}  
    \caption{Here is a schematic diagram of a latent space representation of a quantum circuits generated by UCCSD, which merely depicts a high-dimensional space where the coordinate axes carry no specific meanings. Each circle denotes the operator pool associated with a specific molecule, and each square denotes a specific quantum operator.}  
    \label{fig:embed}  
\end{figure}

For each quantum operator in the source molecule's pool, we identify the most similar operator in the target molecule's pool using the ratio of matching sequence elements. This mapping establishes a bridge between the learned patterns in the source molecule and their potential applications in the target molecule. As illustrated in Fig.~\ref{fig:embed}, we embed quantum operators obtained from UCCSD decompositions across different molecules into a coordinate system, thereby defining a quantum-operator space. Each small square corresponds to a specific operator, with color encoding operator identity. Color proximity indicates greater pairwise similarity. For each molecule, the UCCSD-derived set of operators constitutes an operator pool, depicted as a circumscribed circle that encloses the corresponding squares. Distinct circles therefore represent different pools. This visualization supports comparative analysis of operator composition across molecules through the spatial distribution and color similarity within and between pools. 



\begin{figure}[htbp]
    \centering
    \includegraphics[width=0.8\linewidth]{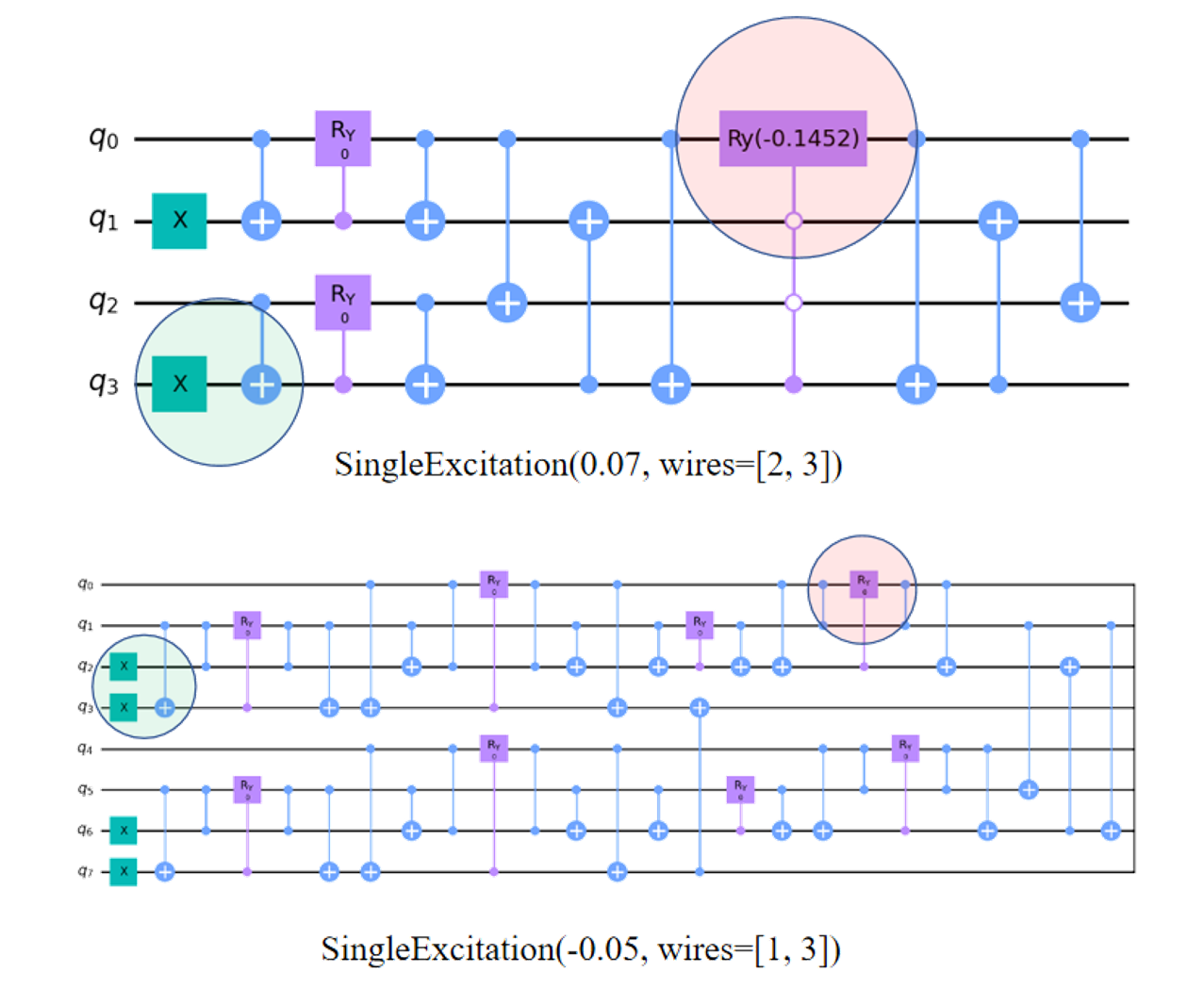}  
    \caption{\texttt{qml.SingleExcitation(0.07, wires=[2, 3])} and \texttt{qml.SingleExcitation(-0.05, wires=[1, 3])} exhibit similarity and they are likewise similar as components with the UCCSD-decomposed circuits of different molecules.}  
    \label{fig:similar}  
\end{figure}

For similarity measure, we establish correspondences between quantum operators. Each operator is encoded as a string and compared using Levenshtein Distance based sequence-matching algorithm~\cite{yujian2007normalized}, which computes similarity by identifying maximally matching blocks and recursively accounting for substitutions.  The resulting score quantifies the pairwise similarity between operator encodings. As shown in Fig.~\ref{fig:similar},  if distinct quantum operators share a certain degree of similarity, they exhibit corresponding similarity across different circuits.
While our current implementation uses this straightforward similarity measure, the framework allows for more sophisticated approaches.

\subsection{Transfer Learning Pipeline}
\begin{figure*}[h]
    \centering
    \includegraphics[width=0.9\textwidth]{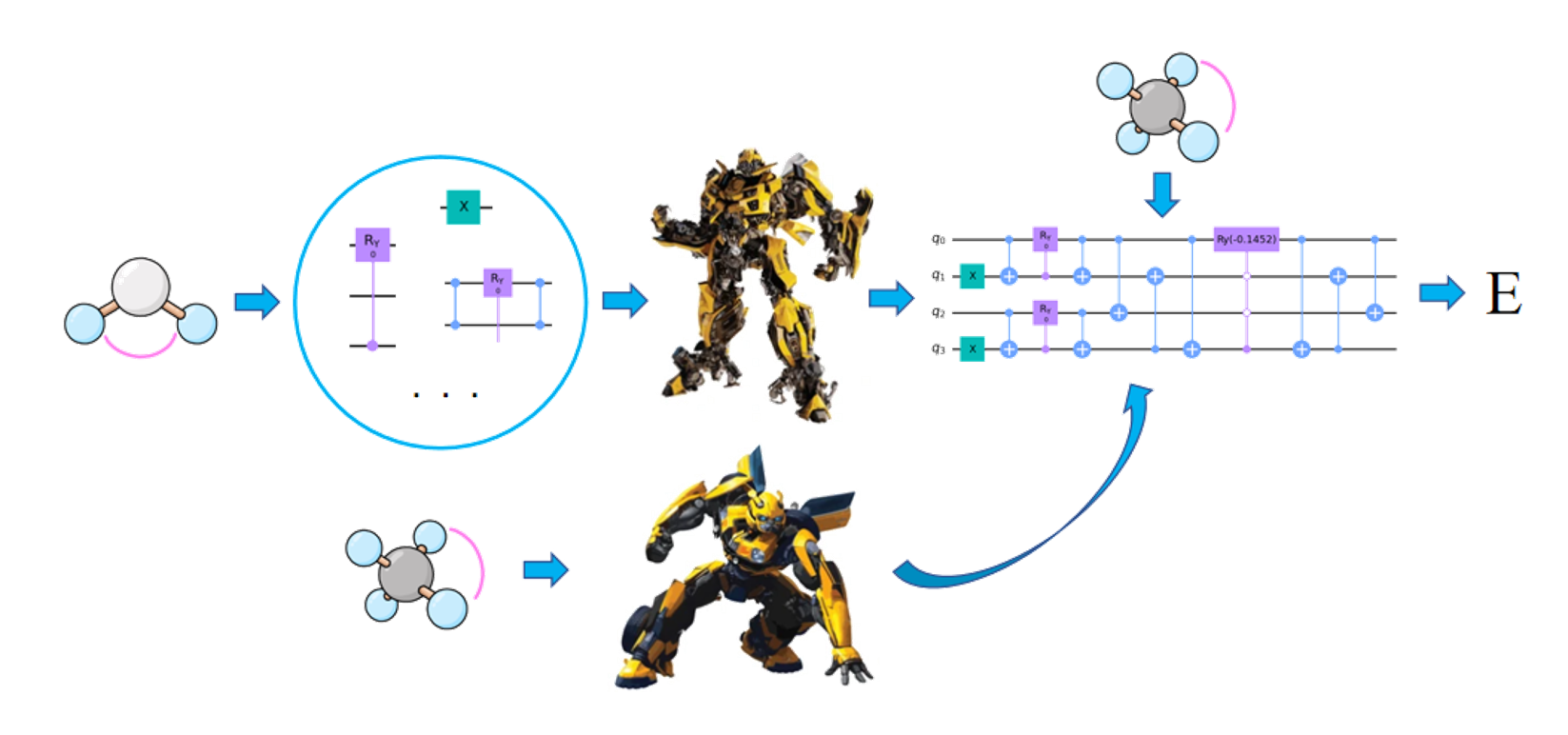}
    \caption{Schematic demonstration for the transfer learning pipeline. The Transformer figures in the first row represent the original transformer model, whereas those in the second row represent the fine-tuned transformer model. \textbf{E} denotes the ground state energy.}
    \label{fig:pipeline}
\end{figure*}

Our complete methodology follows the pipeline below. Fig.\ref{fig:pipeline} provides a visual illustration of this process:

\begin{enumerate}
\item \textbf{Source Molecule Training:} We first train a Generative Quantum Eigensolver (GQE) model on a source molecule (e.g., \ce{H2}). The backbone of GQE is a transformer model. This model learns to generate sequences of quantum operators that progressively reduce the energy expectation of the source molecule's Hamiltonian.
\item \textbf{Model Loading:} The trained model, which encapsulates patterns of effective quantum operators for the source molecule is loaded for transfer to the target molecule.
\item \textbf{Model Structure Adaptation:} We adapt the model architecture to accommodate differences in operator pool size between source and target molecules. This involves adjusting the dimensionality of embedding layers and output heads while preserving the core transformer architecture that has captured generalizable patterns in quantum operator sequences.
\item \textbf{Operator Mapping Establishment:} Using our similarity measurement framework, we establish mappings between source and target molecule operators. This mapping preserves the functional relationships between operators while accounting for variations in qubit dimensionality.
\item \textbf{Selective Fine-Tuning (Optional):} When additional computational resources are available, we selectively fine-tune key model components (embedding layers and output heads) using new data from the target molecule. This step bridges remaining gaps between source and target domains while maintaining most of the transferred knowledge.
\item \textbf{Target Molecule Energy Prediction:} The adapted model generates quantum operation sequences for the target molecule. These sequences are evaluated against the target molecule's Hamiltonian to predict ground state energies.
\end{enumerate}

This pipeline enables efficient knowledge transfer across molecular systems, reducing the computational resources required for accurate ground state energy predictions. By leveraging the similarities in quantum operator patterns, our approach circumvents the need to train models from scratch for each new molecule, while maintaining prediction accuracy.

The entire framework is designed with extensibility in mind and allows progressive improvements. Even with our baseline implementation using simple string similarity and minimal fine-tuning, we observe benefits that establish a baseline on the potential of quantum operators representation transfer.


\section{Results and Discussion}

Our experimental framework utilizes molecular systems provided by PennyLane Molecules~\cite{Utkarsh2023Chemistry}, which constitute ideal benchmark candidates due to comprehensive  calculations.

\subsection{Directional Asymmetry in Quantum Operator Transfer}

Our initial investigations reveal a clear transfer effect between molecular systems, with a directional asymmetry that provides insight into the nature of quantum operator representation learning. When comparing bidirectional transfer between simple and complex molecular(relative size) systems, we consistently observe lower prediction errors when transferring from complex to simple molecules compared to the reverse direction.

Using \ce{H2} as our reference simple molecule, we conducted bidirectional transfer experiments with several larger molecular systems. Table~\ref{first-table} shows this asymmetry by comparing Root Mean Square Error (RMSE) and absolute error in ground state energy predictions. In all tested molecular pairs, the prediction error when using the simple molecule as source and transferring to more complex targets was significantly higher than when those same complex molecules served as the source.

This asymmetry can be attributed to the representation capacity of the quantum operators. The quantum operators for simple molecules like \ce{H2} encode a limited set of electronic structures and interactions, providing insufficient representation capacity to capture the richer electronic behaviors in larger molecules. Conversely, the operators of complex molecules contain a more comprehensive set of quantum operations that can effectively subsume the behaviors needed for simpler molecules, enabling more successful knowledge transfer in that direction.

\begin{table}[t!]
\begin{tabular}{lllll}
\hline
$Source \rightarrow Target$  & GT & Predicted & Abs Error & RMSE\\ \hline \hline
$\ce{H2} \rightarrow \ce{LiH}$      & -7.8726 & 10.7570 & 18.6297  & 18.6190   \\
$\ce{LiH} \rightarrow \ce{H2}$    & -1.1262  & 2.6323  & \textbf{3.7585}  & \textbf{3.7513} \\
$\ce{H2} \rightarrow \ce{HF}$    & -98.6204   & 11.4031  & 110.0235 & 109.9723 \\ 
$\ce{HF} \rightarrow \ce{H2}$    & -1.1276  & -73.1013  & \textbf{71.9737}  & \textbf{71.9833} \\
$\ce{H2} \rightarrow \ce{He2}$   & -5.7530  & 9.3976 & 15.1506  & 15.1098 \\ 
$\ce{He2} \rightarrow \ce{H2}$   & -1.1254  & -5.1140 & \textbf{3.9886}  & \textbf{4.0099}  \\
 \hline
\end{tabular}
\caption{\label{first-table} All tabulated energies are reported in hartree (Ha), where GT corresponds to the energy obtained through VQE. }
\end{table}

\subsection{Transfer Performance Analysis with $\ce{H2}$ as Source Molecule}

To systematically evaluate our approach, we conducted transfer experiments using \ce{H2} as the source molecule and a diverse set of target molecules including various hydrogen chains (\ce{H4}, \ce{H5}, etc.) and more complex molecular systems. Table~\ref{second-table} summarizes the prediction errors across these transfers, measured by RMSE and absolute error metrics.

Our results indicate that transfer performance varies significantly based on structural similarity between source and target molecules. Hydrogen chains (\ce{H4}, \ce{H5}, etc.) demonstrated relatively valid transfer trends with metrics values consistently, suggesting that our representation framework effectively captures similarities in linear hydrogen systems despite their different sizes.

However, performance degraded substantially for molecules with different elemental compositions and electronic structures. Molecules like \ce{LiH}, \ce{BeH2}, and \ce{He2} exhibited larger errors, indicating limited transfer efficacy. This performance limitation likely stems from our current naive baseline representation approach, which relies on simple string similarity without incorporating deeper chemical or quantum mechanical principles.

The performance pattern confirms our hypothesis that even simple text representation of quantum operators contains transferable information, while simultaneously highlighting the need for more sophisticated representation methods that can better capture the quantum mechanical equivalences between different molecular systems.

\begin{table}[t!]
\begin{tabular}{lllll}
\hline
$\ce{H2} \rightarrow Target$  & GT & Predicted & Abs Error & RMSE\\ \hline \hline
$\ce{H3+}$      & -1.2508 & 6.8022 & 8.0530  & 8.0517   \\
$\ce{H4}$    & -2.1459  & 7.6440  & 9.7899  & 9.7558 \\
$\ce{H5}$    & -2.5978   & 10.0606  & 12.6585 & 12.6338 \\ 
$\ce{H6}$    & -3.1936  & 9.6137  & 12.8073  & 12.7677 \\
$\ce{H7}$   & -3.6535 & 11.0980 & 14.7515  & 14.7154 \\ 
$\ce{He2}$   & -5.7530  & 9.3976 & 15.1506  & 15.1098  \\
$\ce{HeH+}$   & -2.8487  & -1.0765 & 1.7721  & 1.7826 \\ 
$\ce{LiH}$   & -7.8726  & 10.7570 & 18.6296  & 18.6189 \\ 
$\ce{BeH2}$   & -15.5898  & 10.9515 & 26.4913  & 26.4601 \\ 
 \hline
\end{tabular}
\caption{\label{second-table} Transfer performance analysis with $\ce{H2}$ as source molecule, where GT corresponds to the energy obtained through VQE. }
\end{table}

\subsection{Unexpected Effects of Fine-Tuning}

Contrary to conventional wisdom in transfer learning, our fine-tuning experiments yield some unexpected results of performance degradation rather than improvement. Table~\ref{third-table} shows this counter-intuitive finding, showing how fine-tuning the embedding and output layers using target molecule data actually increased prediction errors across multiple molecular targets.
This phenomenon can be explained by the limited expressivity of our source molecule and the differences between source molecule and target molecule. In Table~\ref{third-table}, we use \ce{LiH} as the source. The pre-trained model has learned patterns from an extremely restricted operator space that lacks the representational complexity needed for larger molecules. In this scenario, fine-tuning appears to destabilize the already limited learned representations without providing sufficient additional information to improve predictions.

The fine-tuning process likely causes catastrophic forgetting of the limited but useful patterns learned from the source molecule, while the restricted fine-tuning dataset provides insufficient information to develop effective representations for the target molecule. This also explains why fine-tuning yields valid performance increase for both \ce{H6} and \ce{H2} systems. It reinforces the importance of developing more expressive baseline representations that can better bridge the gap between molecules of different complexities.



\begin{table}[htbp]
\centering

\begin{tabularx}{\linewidth}{l X X X} 
\toprule
Molecules  & RMSE(without fine-tune) & RMSE(with fine-tune)\\
\midrule
$\ce{HF}$      & 102.1811 & \textbf{119.8051}  \\
$\ce{H6}$    & \textbf{14.5846}  & 13.1104  \\
$\ce{NeH+}$    & 130.4021   & \textbf{149.3346}   \\ 
$\ce{OH-}$    & 77.6657  & \textbf{118.1038}   \\
$\ce{H2}$   & \textbf{3.7513} & 0.1634  \\ 
\bottomrule
\end{tabularx}
\caption{\label{third-table} Unexpected effects of fine-tuning while \ce{H6} and \ce{H2} yields valid performance increase.}
\end{table}

\subsection{Computational Efficiency}

A primary motivation for our transfer learning approach is to reduce the computational resources required for molecular ground state energy calculations. Table~\ref{forth-table} compares the computational requirements of our transfer method against training new GQE models from scratch for each target molecule. Our transfer learning pipeline demonstrates remarkable efficiency advantages. Taking small molecules \ce{H2} and \ce{LiH} as examples, training a new molecule GQE model from scratch requiring 5000 epochs to achieve convergence and obtain stable models, with training times of 3692 seconds and 3810 seconds respectively. Our transfer learning strategy demonstrates a $100\times$ speedup compared to training from scratch.

This efficiency stems from eliminating the need to generate extensive training datasets through repeated quantum circuit evaluations, which constitute the most resource-intensive component of GQE training. This efficiency gain becomes increasingly significant for larger molecular systems, where quantum circuit evaluation costs grow polynomially with system size. The ability to bypass most of these evaluations through knowledge transfer represents a substantial advancement toward practical quantum-classical algorithms for real-world molecular systems.


\begin{table}[htbp]
\centering

\begin{tabularx}{\linewidth}{l X X X} 
\toprule
Molecules  & Time(without fine-tune) &  Time(with fine-tune)\\
\midrule
$\ce{HF}$      & 344 & 751  \\
$\ce{H6}$    & 658  & 1323  \\
$\ce{NeH+}$    & 321   & 699   \\ 
$\ce{OH-}$    & 233  & 701   \\
$\ce{H2}$   & 20 & 114  \\ 
\bottomrule
\end{tabularx}
\caption{\label{forth-table} Energy prediction times (in seconds) for various molecules. The operator sequence length here is 200, with fine-tuning performed over 200 epochs.}
\end{table}

\section{Conclusion}

In this work, we introduced a novel representation transfer learning framework for quantum operators. Our approach treats UCCSD quantum operators as strings and establishing similarity-based mappings between molecules within GQE paradigm. We demonstrated that quantum circuit knowledge can be effectively transferred from the source to target molecules, significantly reducing computational requirements while maintaining performance for similar molecular structures.

These initial findings were achieved with a naive baseline setting on quantum operator representation, similarity calculation, model complexity and training procedures, establishing a proof-of-concept demonstration rather than an optimized solution. This conservative approach underscores the significant potential of our quantum operator representation transfer framework. More rigorously, quantum circuits exhibit inherent temporal ordering, and the granularity of SMILES-like representation used in this work remains limited in expressivity. The enhancements are expected to significantly improve performance metrics in the optimized version. 


%


\bibliographystyle{IEEEtranS}
\bibliography{main}

\end{document}